\documentclass[aip,jcp,reprint,longbibliography]{revtex4-1}

\usepackage[usenames,dvipsnames]{color}
\usepackage{amssymb}
\usepackage{amsmath}
\usepackage{hyperref}
\usepackage{graphicx}
\usepackage{subfigure}
\usepackage{xspace}

\newcommand{\DG}{\Delta G}
\newcommand{\DGcnf}{\Delta G_\mathrm{cnf}}

\newcommand{\di}{{\mathrm{d}}}

\date{\today}

\begin{document}

\title{
A new configurational bias scheme for sampling supramolecular structures
}
\author{Robin De Gernier,$^{1}$ Tine Curk,$^{2}$ Galina V.\ Dubacheva,$^{3}$  
Ralf P.\ Richter,$^{3,4,5,6}$ and Bortolo M.\ Mognetti
\cite{corrauthor}$^{,}$
}
\address{
$^{1}$Center for Nonlinear Phenomena and Complex Systems, Universit\'e Libre de Bruxelles, Code Postal 231, Campus Plaine, B-1050 Brussels, Belgium \\
$^{2}$Department of Chemistry, University of Cambridge, Cambridge CB2 1EW, United Kingdom\\
$^{3}$Biosurfaces Unit, CIC biomaGUNE, Paseo Miramon 182, 20009 Donostia - San Sebastian, Spain\\
$^{4}$Universit\'e Grenoble Alpes, DCM, 38000 Grenoble, France
\\
$^{5}$CNRS, DCM, 38000 Grenoble, France
\\
$^{6}$Max Planck Institute for Intelligent Systems, 70569 Stuttgart, Germany
}

\begin{abstract}
We present a new simulation scheme which allows an efficient sampling of reconfigurable supramolecular structures made of polymeric constructs functionalized by reactive binding sites. The algorithm is based on the configurational bias scheme of Siepmann and Frenkel and is powered by the possibility of changing the topology of the supramolecular network by a non--local Monte Carlo algorithm. Such plan is accomplished by a multi--scale modelling that merges coarse-grained simulations, describing the typical polymer conformations, with experimental results accounting for free energy terms involved in the reactions of the active sites. We test the new algorithm for a system of DNA coated colloids for which we compute the hybridisation free energy cost associated to the binding of tethered single stranded DNAs terminated by short sequences of complementary nucleotides. In order to demonstrate the versatility of our method, we also consider polymers functionalized by receptors that bind a surface decorated by ligands. In particular we compute the density of states of adsorbed polymers as a function of the number of ligand--receptor complexes formed. Such a quantity can be used to study the conformational properties of adsorbed polymers useful when engineering adsorption with tailored properties. We successfully compare the results with the predictions of a mean field theory. We believe that the proposed method will be a useful tool to investigate supramolecular structures resulting from direct interactions between functionalized polymers  for which efficient numerical methodologies of investigation are still lacking. 
\end{abstract}

\maketitle


\section{Introduction}



Polymeric constructs functionalized by active groups that can selectively react with complementary groups { are} at the core of many biological systems (e.g.\ cell signaling and protein docking) and are becoming a very popular tool to engineer new functional materials in the field of nanotechnology.\cite{Mirkin_Nature_1996,Alivisatos,Di-Michele_rev,Galina,Kiessling_rev} 
For instance, DNA strands tipped by reactive sequences of single stranded (ss)DNA are currently used to mediate direct interactions between colloids,\cite{Mirkin_Nature_1996,Alivisatos,Di-Michele_rev} in DNA origami to assist the assembly of DNA tiles into complex patterns,\cite{Tile_Origami} or to design supramolecular gels.\cite{Biffi_PNAS,Schmid_PRL}
Functionalized polymers are also used in nanomedicine, in particular in drug 
delivery to engineer selective targeting.\cite{Galina,Francisco_PNAS_2011,Kiessling_rev}

%

Functionalized polymers are difficult to model because their properties result from a synergistic 
effect between the reaction free energy of the functional groups and the polymer conformations 
that are sharply constrained by the tight binding acting between 
 reacting spots.\cite{Dreyfus_2010,Mirjam_JCP_2011,SoftMatter_2012} These two contributions to the free energy are comparable (though usually of opposite sign) and, in the interesting regimes, are accessible by thermal fluctuations.\cite{Mirjam_JCP_2011} Hence a  statistical mechanics treatment of these systems needs to account for these two effects.\cite{PNAS_2012,Patrick_JCP_2012,Stefano_JCP_2013}

This { leads} to a multi--scale problem that hampers the modelling of these systems. In particular an adequate description of the reactive binding sites requires atomistic models that become unpractical when sampling polymer conformations. 
This can be better explained by considering the two systems that will be treated in this paper. First we will study the hybridisation of tethered inert strands of ssDNA terminated by a reactive sequence as used in DNA coated colloids (DNACCs). 
{
Here a detailed model necessary to properly describe the hybridisation free energy of the reactive sequences\cite{OXDNA} cannot be employed, in realistic computational time, to study typical DNACCs made of thousand of different ssDNA (long of up to 50 base pairs) terminated by short strings of active bases.\cite{Rogers_PNAS}
}
In a second system we will study the conformation and the density of states of functionalized polymers
adsorbed by ligands distributed on a surface as motivated by recent experiments.\cite{Galina} 
Similarly to the case of DNACCs, a proper sampling requires exploring many 
{ configurations grouped in different}
 topologies in which different receptors bind different ligands. Realistic dynamics of atomistic models cannot access the timescales of 
such systems.


{ In this paper we study the possibility of designing non--local Monte Carlo (MC) moves to sample between supramolecular polymer configurations with different topologies.} Specifically we propose an algorithm that could, in one { step}, bind/unbind two tethered strands in a system 
featuring DNA-mediated interactions
 or that could attach/detach a receptor of a functionalized polymer to/from a ligand tethered to a surface. In doing so we will employ
a multi--scale approach in which the free energy of the reacting sites is taken into account implicitly,
using accessible experimental results.

%

Some steps in this direction have already been taken.\cite{PNAS_2012,Patrick_JCP_2012} In particular in Ref.\ \onlinecite{PNAS_2012} we used MC Rosenbluth Sampling \cite{DF_book} to estimate the hybridisation free energy of tethered single stranded DNA constructs. 
{
 Because Rosenbluth weights can be linked to the free energy of the constructs,
\cite{DF_book} it is possible to calculate the configurational part of the hybridisation free energy by 
comparing independent Rosenbluth simulations of free and hybridised strands.
}
%
%

{ 
In this paper we want to extend these methods to a dynamic algorithm in which the supramolecular network is reconfigured on the fly.
}
There are different reasons for { aiming at} this step. Notoriously the quality of the sampling in 
{ static } Rosenbluth simulations  becomes poor for long polymeric constructs.\cite{Mooij,Batoulis} 
Moreover, from a more practical point of view, a dynamic scheme is much more versatile because it allows to study a broader range of systems for which pre--computing Rosenbluth weights (as done in Refs.\ \onlinecite{PNAS_2012,Patrick_JCP_2012}) would be unfeasible. In particular this has motivated the study of the targeting problem presented in the second part of the paper.


This paper is organised as follows. 
In Sec.\ \ref{Sec:TheMethod} we present our algorithm. We present the
 multi--scale approach, detail the algorithm by which supramolecular structures are generated, and 
 derive the acceptance rules used to swap between them. In view of the similarity with the configurational bias MC (CBMC) scheme,\cite{CBMC} and of its ability to { swap between configurations with different topologies, we label the new algorithm topological CBMC (tCBMC). }
In Sec.\ \ref{Sec:DNACC} we test tCBMC for DNA--coated colloids systems. In particular we show how tCBMC can measure the hybridisation free energy of two tethered constructs in agreement with previous studies.\cite{PNAS_2012,Patrick_JCP_2012}
In Sec.\ \ref{Sec:SE} we then consider a polymer functionalized by receptors targeting ligands distributed on a surface. In particular we show how tCBMC, in tandem with a powerful umbrella sampling scheme,\cite{SUS} allows to compute the density of states of an adsorbed polymer as a function of the number of functionalized ligands.
 We validate our finding using a mean field (MF) theory that we present in appendix \ref{app:MF}. 
 Finally in Sec.\ \ref{Sec:Dis} we present the conclusions and the perspectives of our work.





\begin{figure}
\begin{center}
\vspace{-2.cm}
\includegraphics[scale=0.28]{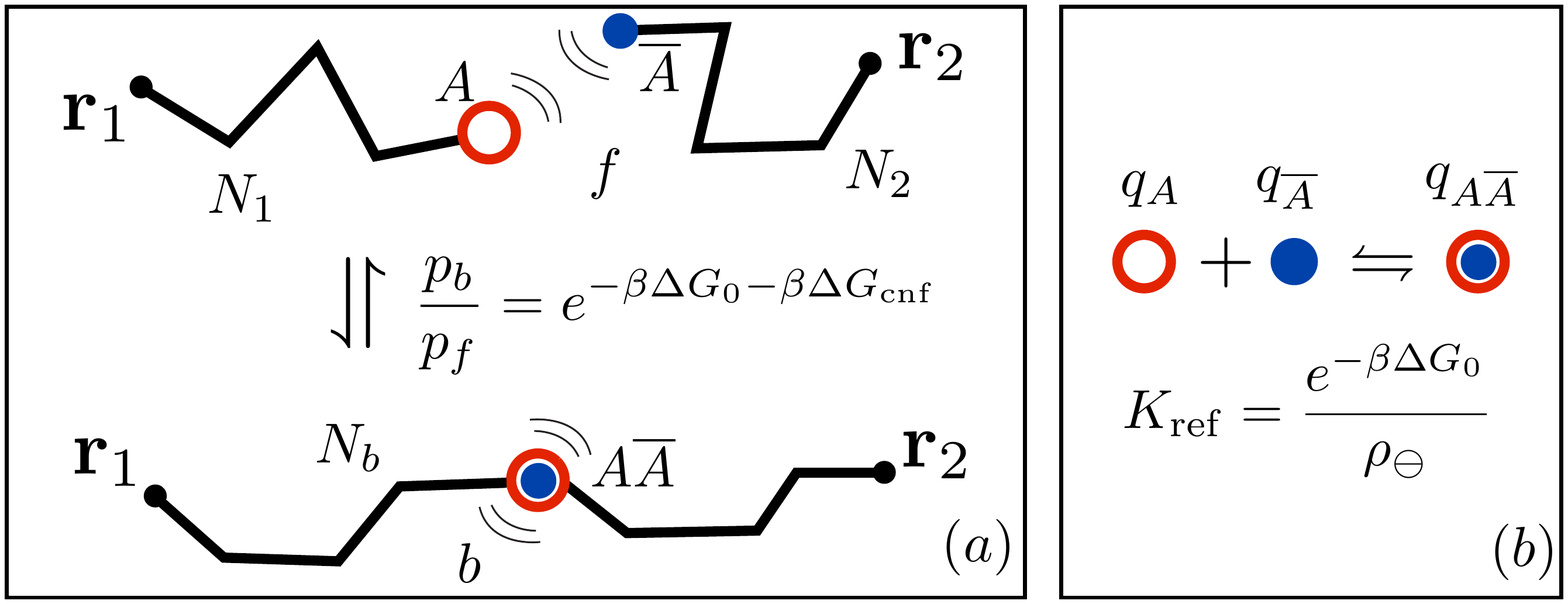}
\end{center}
\vspace{-2.cm}
\caption{\label{fig1} $(a)$  The hybridisation of two tethers functionalized by reactive sites is controlled by the hybridisation free energy of the free active groups in solutions $\DG_0$ $(b)$, and by a configurational term $\DGcnf$ due to the tethering constraint.}
\end{figure}

\section{The method}\label{Sec:TheMethod}





Fig.\ \ref{fig1}$a$ depicts the typical situation that we are interested in modelling. Two polymeric constructs, tethered in ${\bf r}_1$ and ${\bf r}_2$, are tipped by complementary reactive elements ($A$ and $\overline A$). 
We discern between two possible states in which the constructs are either free ($f$) or react giving rise to a bound  state ($b$). In the latter case we say that a supramolecular structure, spanning between ${\bf r}_1$ and ${\bf r}_2$, has been formed. 
{
Here we introduce an algorithm that samples directly between configurations of type $f$ and 
of type $b$. 
}
For simplicity we keep the tethering points ${\bf r}_1$ and ${\bf r}_2$ fixed (this constraint can be removed combining the current technique with standard local algorithms).

Our coarse-grained approach does not model the atomistic details of the reactive elements. 
Instead we use implicit terms ($q_A$, $q_{\overline A}$ and $q_{A\overline A}$) as internal partition function of the { active groups and bound complex} ($A$, $\overline A$ and $A\overline A$, respectively).
The internal partition functions can be linked to the equilibrium constant of the dimerisation reaction between free reacting groups in solution (Fig.\ \ref{fig1}$b$).
In particular the equilibrium condition between the chemical potential of the reactants
$\mu_b = \mu_{f1} + \mu_{f2}$ implies the following relation\cite{Mirjam_JCP_2011,SoftMatter_2012}
\begin{eqnarray}
{q_{A\overline A} \over q_{A} q_{\overline A}} = K_\mathrm{ref} = {e^{-\beta \Delta G_0} \over \rho_\ominus} \, ,
\end{eqnarray}
where $\rho_\ominus$ is the standard concentration.
The partition functions of the free ($f$) and bound ($b$) tethers can then be { derived summing over 
all the possible constructs' configurations }
\begin{eqnarray}
Z_f &=& q_{A} q_{\overline A} \int \di c_f \, e^{-\beta H^{(f)}(c_f)}  \, ,
\label{eq:Zf}
\\
Z_b &=& q_{A \overline A} \int \di c_b \, e^{-\beta H^{(b)}(c_b)} 
\delta( f_\mathrm{ee}(c_b)-r_{12})
\, ,
\label{eq:Zb}
\end{eqnarray}
where $c_f$ represents two polymers emanating from ${\bf r}_1$  and ${\bf r}_2$, while $c_b$ is a single 
polymer branch (see Fig.\ \ref{fig1}).
In Eq.\ \ref{eq:Zb} $f_\mathrm{ee}(c_b)$ is the function that provides the end-to-end distance of a bound configuration ($c_b$) and we have defined $r_{12}=|{\bf r}_1 - {\bf r}_2|$.
In Eqs.\ \ref{eq:Zf} and \ref{eq:Zb} $H^{(f)}$ and $H^{(b)}$ are the configurational energies of the free and of the bound constructs. 
They also account for the interactions of the chain with the environment (hard walls, other polymers, {\em etc}.). 
In this study the polymeric constructs are modelled by flexible freely-jointed chains (FJCs) made of $N_1$ and $N_2$ segments for the free constructs and $N_b=N_1+N_2$ segments for the bound construct. 
The nature of $H^{(b)}$ and $H^{(f)}$ will be further specified in the next sections.
{
The method of Fig.\ \ref{fig1}$a$ may resemble identity swap MC schemes that have
been used, for instance, to sample populations of polymers with different lengths by removing 
monomers from longer chains and regrowing them at the end terminals of the shorter chains.\cite{SiepMD}
However in Fig.\ \ref{fig1}$a$ we have to account for the loss of three degrees of freedom of the hybridised 
chain due to the fixed--end--point constraints.
} 

Notice that we have used a point--like representation for the reactive elements. This may look limiting, for instance, in the case of DNACC systems  where the length of the hybridised segments can be comparable with the Kuhn length of the constructs. { In this regard,} we observe that it is rather straightforward to generalise our procedure to more detailed models that include a non trivial representation of the reactive groups.

Here we explain how we create new polymer configurations.
Like in Rosenbluth sampling,
in order to generate free configurations $f$, we grow two open chains by sequentially adding $N_1$ and $N_2$ segments ${\bf u}_i$ starting from ${\bf r}_1$ and ${\bf r}_2$ respectively.\cite{optimal_grow}
The $i$--$th$ segment (${\bf u}_i$) is sampled within $k$ possible ones (${\bf u}_{i,\alpha}$, with $\alpha=1,\cdots k$) that are generated with a uniform distribution. 
{ In this work we use $k=20$.}
The segment ${\bf u}_i$ is chosen within the $k$ possibilities with probability 
$p_{i,\alpha} \sim { \exp[-\beta H^{(f)}_i({\bf u}_{i,\alpha})]   }$. 
$H^{(f)}_i({\bf u})$ is the interaction of the segment ${\bf u}$ with the surrounding environment, including the fraction of chains already grown.\cite{DF_book}
If  $W^{(f)}_i = k^{-1}\sum_{\alpha=1}^k \exp[-\beta H^{(f)}_i({\bf u}_{i,\alpha})] $, we can define the Rosenbluth weight of the newly generated configuration ($c_f$) in the standard way $W^{(f)}(c_f)=\prod_{i=1}^{N_b} W^{(f)}_i$. Following the previous procedure, a free configuration $c_f$ is generated with probability
\begin{eqnarray}
p(c_f) &=& {e^{-\beta H^{(f)}(c_f) }\over W^{(f)}(c_f) } \, .
\end{eqnarray}

A similar algorithm can be used to grow bound constructs $b$. However in this case we have 
to further bias the sampling to satisfy the distance constraint on the end points.
 If the $i-1$--$th$ segment terminates in ${\bf x}_{i-1}$, the $k$ possible segments are generated with probability distribution function given by the normalised density of FJCs made of $N_b-i$ segments with end point 
 distance equal to $|{\bf r}_2-{\bf x}_{i-1} + {\bf u}_{i,\alpha}|$.\cite{DF_book}
This probability function is given by $p(|{\bf r}_2-{\bf x}_{i-1} + {\bf u}_{i,\alpha}|;N_b-i)/p(|{\bf r}_2-{\bf x}_{i-1}|;N_b-i+1)$,
where $p$ is the end--to--end distribution function of FJC constructs.\cite{yamakawa1,yamakawa2} 
In practice, the generation of the trial segments ${\bf u}_{i,\alpha}$ is done by a hit--or--miss algorithm that uses the known end--to--end functions $p$.\cite{yamakawa1,yamakawa2} 
We then define $W^{(b)}(c_b)$ as the Rosenbluth weight of the bound construct which is 
computed as for free polymers but using $H^{(b)}$ instead of $H^{(f)}$. 
Notice that the previous procedure produces a hybridised construct $c_b$ with probability
\begin{eqnarray}
p(c_b) &=& \prod_{i=1}^{N_b} {e^{-{\beta}H^{(b)}_i ({\bf u_i})} \over W^{(b)}_i } { p(|{\bf r}_2-{\bf x}_{i-1} + {\bf u}_\alpha|;{N_b}-i) \over p(|{\bf r}_2-{\bf x}_{i-1}|;N_b-i+1) }
\nonumber \\
&=& {\delta (f_{ee}(c_b)-r_{12}) \over p(r_{12}, N_b) }  {e^{-{\beta}H^{(b)} (c_b)} \over W^{(b)} (c_b)} \, ,
\end{eqnarray}
{where we recall that $f_{ee}$ is the end--to--end distance of the bound configuration $c_b$. }

Growing fixed--end chains has already been used in polymer simulations.\cite{Esco_EP,Vendr_EP,Dij_EP,Siep_EP} 
{ Instead}, what we propose here is { an algorithm that samples} between constructs of type $c_b$ and constructs of type $c_f$.
%
This can be done using the following acceptance rules 
\begin{eqnarray}
acc_{b\to f} &=& \mathrm{min} [ 1, {\rho_\ominus \over p(r_{12},N_b)}  { W^{(b)}(c^{(n)}_b) \over e^{-\beta \Delta G_0} W^{(f)}(c^{(o)}_f) } ]
\nonumber 
\\
acc_{f\to b} &=& \mathrm{min} [ 1,  {p(r_{12},N_b)\over \rho_\ominus}  {e^{-\beta \Delta G_0} W^{(f)}(c^{(n)}_f) \over  W^{(b)}(c^{(o)}_b)} ] \, ,
\nonumber 
\\
\label{acceptances}
\end{eqnarray}
as can be derived using Eqs.\ (1-5). In the previous equations $c^{(n)}$ and $c^{(o)}$ distinguish the trial (new) configuration from the current (old) system configuration. As in CBMC\cite{CBMC} the Rosenbluth weight of the old configuration is computed by re--growing the chains.
{It is important to notice that our method is not constrained to the knowledge of the end-to-end distance functions $p$. In particular using the self-adapting fixed-end-point scheme of Ref.\ \onlinecite{Siep_EP}, it is possible to design guiding probability distributions that can replace $p$ when directing the growth of the chains toward the target point. }

Finally using the acceptance rules (Eqs.\ \ref{acceptances}) it is possible to calculate the ``polymeric free energy'' associated to the formation of the { bound construct} as
\begin{eqnarray}
\DG_\mathrm{hyb} = \DG_0 + \DGcnf = -k_B T\log \Big[ {L_b \over L_f} \Big] \, ,
\label{eq:DG}
\end{eqnarray}
where {$L_b$} and {$L_f$} are the number of times that the simulation has visited a bound and a free state.


\begin{figure}
\begin{center}
\includegraphics[scale=0.2]{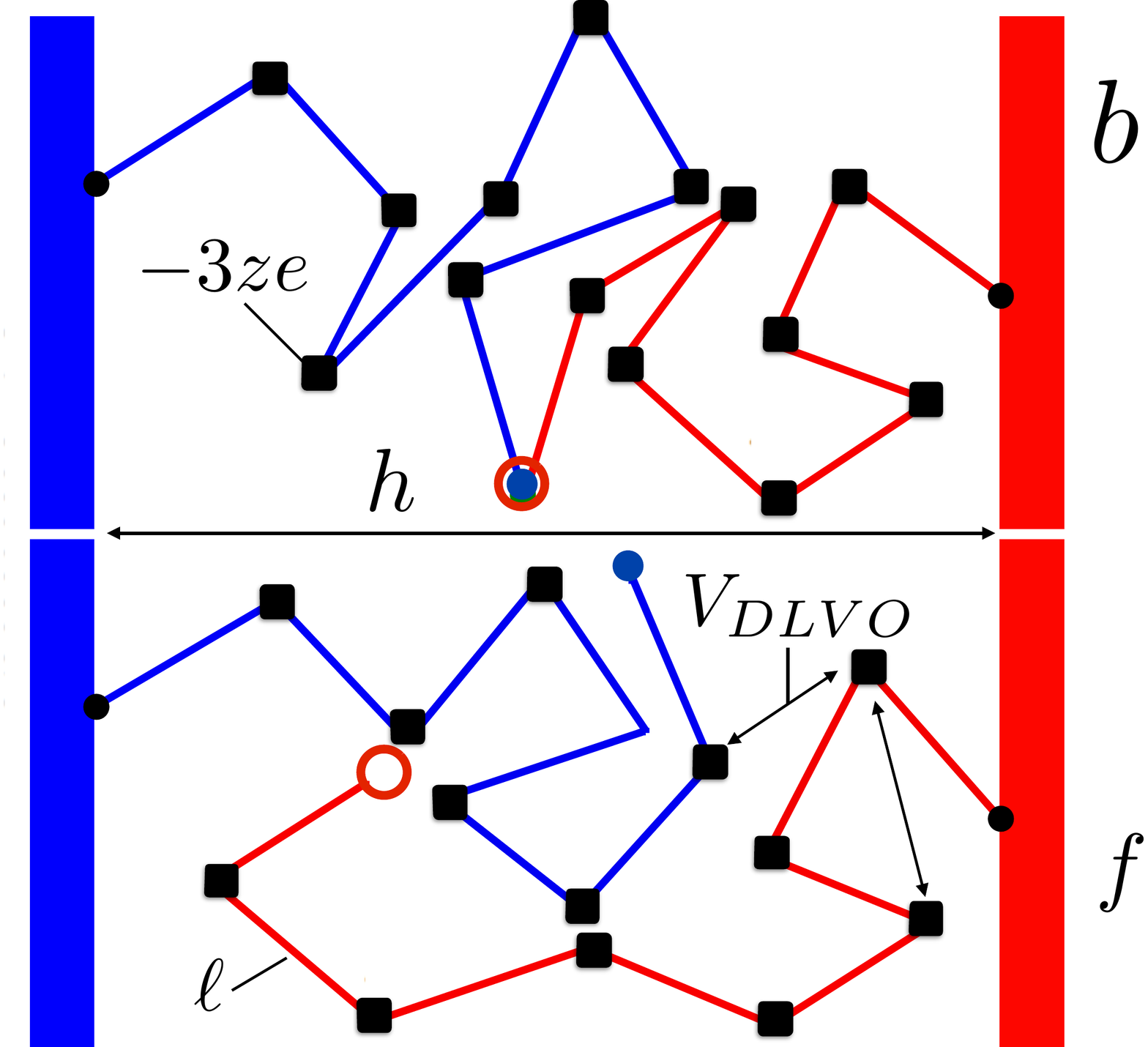}
\end{center}
\caption{\label{figDNA} 
Coarse-Grained model of DNA strands tipped by reactive sites (circles). The squares represent the charges of the negative backbone that are coarsened (in groups of three) at the junction points of the chains and rescaled by a normalisation factor ($z$) that is provided by DLVO theory.\cite{HMcDbook}
}
\end{figure}

\section{DNA-Coated Colloids}\label{Sec:DNACC}

In this section we utilise tCBMC to compute the 
hybridisation free energy associated to bridge formation in DNACC systems (Fig.\ \ref{figDNA}) and compare the results with previous attempts.\cite{PNAS_2012,Patrick_JCP_2012}
We map a single--stranded DNA into a FJC \cite{ZhangEl} with unit length segment equal to $\ell=1.25\,$nm. 
The unit length segment $\ell$ has been chosen comparing end-to-end distances with a more 
accurate model of the DNA strands.\cite{OXDNA}
Each segment represents three bases, resulting in an averaged distance between nucleotides compatible with the experimental results \cite{ssDNA_rig1,ssDNA_rig2} 
(0.43-0.5$\,$nm). The negative charges of the backbones are then grouped at the junction point 
between two unit segments (Fig.\ \ref{figDNA}).
A similar model was used in Ref.\ \onlinecite{Mladek}. 
We consider the tethers studied by Rogers {\em et al.}\ \cite{Rogers_PNAS} of a Poly(T) 
string of 65 nucleotides terminated by a reactive sequence. The interaction between two charges 
($i$ and $j$) placed at distance $r_{ij}$ is provided by a screened potential as given by the  
DLVO theory\cite{HMcDbook}
\begin{eqnarray}
V_\mathrm{DLVO}(r_{ij}) = {(z e)^2\over 4 \pi \epsilon_0 \epsilon_R} {\exp[-r_{ij}/\lambda_D] \over r_{ij} } \,
\label{eq:DLVO}
\end{eqnarray}
where $z$ is the effective charge correction\cite{HMcDbook}
\begin{eqnarray}
z={\exp[a/\lambda_D] \over 1+a/\lambda_D } \, .
\label{eq:z}
\end{eqnarray}
In Eq.\ \ref{eq:DLVO} $\lambda_D$ is the Debye length,\cite{HMcDbook}
and $\epsilon_0$ is the vacuum dielectric constant. 
{ In the implicit solvent representation of Eq.\ \ref{eq:z}}, $a$ is the excluded salt region which has been taken equal to $a=0.5\,$ nm.\cite{ZhangEl} 
The temperature is fixed at {$T=308\,$K},\cite{HMcDbook} $\epsilon_R=75$, and we used a monovalent salt concentration equal to 125$\,$mM.\cite{Rogers_PNAS}
The free constructs are made of $N_1=N_2=$21 segments, while the hybridised chain is made of $N_b=N_1+N_2$ segments with fixed end--points. 
$H^{(b)}$ and $H^{(f)}$ are then given by the sum over all the pairs of charges of the DLVO interaction
(Eq.\ \ref{eq:DLVO}) augmented by the impermeable wall term. 
Notice that the end--points of the free chains also carry a charge and that the charge on the middle point of the bound construct is doubled. 
The excluded wall term constrains the constructs to remain within two parallel planes placed at distance $h$ (Fig.\ \ref{figDNA}). 
Below, we consider the case in which the two tethering points are positioned opposite each other (i.e. on a line perpendicular to the two surface planes) and do not move along the surface.
%


Simulations are developed following the scheme presented in the previous sections. In each MC cycle, we either attempt to change topology (with 20\% probability) or we implement a ``standard'' MC move (with 80\% 
 probability). 
The change-of-topology movement attempts to hybridise or to free a bound state with equal probability. 
When attempting to make (open) a bridge if the state is in the bound (free) state the MC move is rejected. 
``Standard'' MC moves consist either in regrowing  full chains using CBMC (with 20\% probability) or in local rotations of chain branches by mean of pivot and double pivot MC moves.
The swap between free and bound states is done as described in the previous section using the acceptance rules given by Eq.\ \ref{acceptances}. The hybridisation free energy is then computed using Eq.\ \ref{eq:DG} by counting the number of times that bound and free states have been visited.

It is convenient to bias the run to explore a comparable number of bound and free configurations. 
This can be done by using a free energy bias $\Delta G_\mathrm{bias}$ that pushes the simulation, say, toward bound states. 
On the fly we can then iteratively correct $\Delta G_\mathrm{bias}$ by a factor $\log({\tilde L_b/\tilde L_f})$ (where {$\tilde L_{b/f}$} is the number of times that the Markov chain has visited the bound/free state starting from the last time that $\Delta G_\mathrm{bias}$ has been corrected) until convergence where $\Delta G_\mathrm{bias} = \DG_\mathrm{hyb}$.

\begin{figure}
\begin{center}
\includegraphics[scale=0.4]{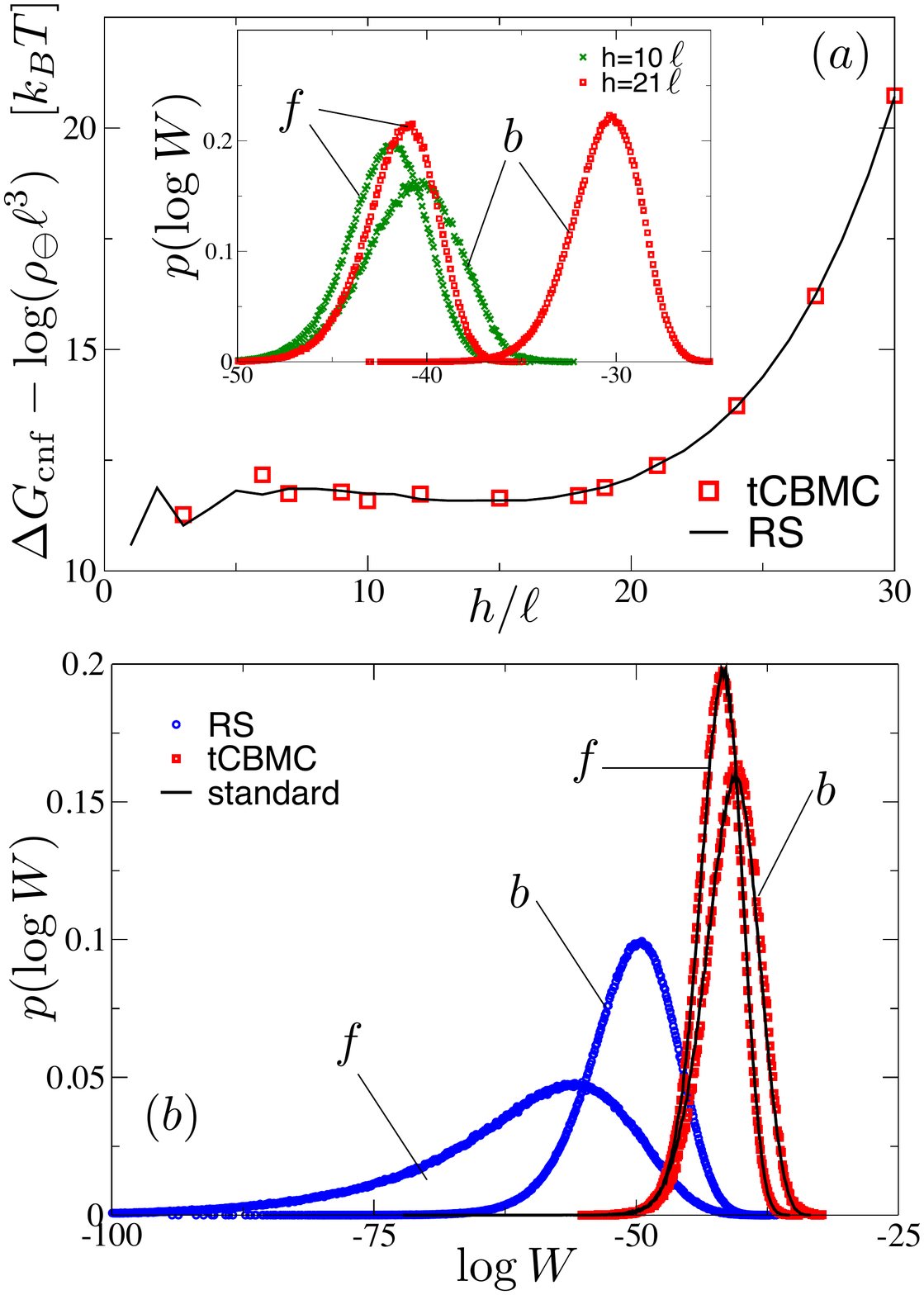}
\end{center}
\vspace{-1.cm}
\caption{\label{figDNARes} ($a$) $\DGcnf$ as a function of the plane distance calculated using Rosenbluth runs\cite{PNAS_2012,Patrick_JCP_2012} (full lines) and using the new proposed algorithm (red squares). 
Inset: distribution functions of the Rosenbluth weights generated in the tCBMC runs for $h/\ell=10$ and $h/\ell=21$.  
($b$) Distribution functions of the Rosenbluth weight for $h=10\ell$ as obtained in Rosenbluth sampling (RS), in tCBMF, and in ``standard'' MC moves.
In part $(b)$ and in the inset of part $(a)$, $f$ and $b$ label free and bound constructs.
}
\end{figure}

Fig.\ \ref{figDNARes}$a$  (symbols) shows the results for $\DGcnf$ (Eq.\ \ref{eq:DG}) at different 
plane-to-plane distances $h$ (Fig.\ \ref{figDNA}). 
Notice that the configurational free energy cost $\DGcnf$ has been translated 
by $k_B T \log (\rho_\ominus \ell^3)$.
This factor appears (along with $\DG_0$) as a pre--factor in  
the acceptance rules in Eq.\ \ref{acceptances} (notice that\cite{yamakawa1,yamakawa2} $p\sim \ell ^{-3}$). 
If compared with the static Rosenbluth method of Refs.\ \onlinecite{PNAS_2012,Patrick_JCP_2012} (full lines) the agreement is perfect (within the scattering due to the noise). 
In the inset of Fig.\ \ref{figDNARes}$a$ we report the probability distribution function of the Rosenbluth weight ($W$) of $f$ and $b$ constructs recorded using tCBMC.
While for $h=10\ell$ we have overlap between the free and the bound distributions, for $h=21\ell$ the overlap region is minimal. Nevertheless convergence is achieved also in the latter case. 
This proves the robustness of the method and highlights the importance of using a bias ($\DG_\mathrm{bias}$) to record a sufficiently high number of jumps between $f$ and $b$ states.

To better highlight the different nature of the proposed method with respect to previous studies,\cite{PNAS_2012,Patrick_JCP_2012} 
in Fig.\ \ref{figDNARes}$b$ we compare the Rosenbluth weights distributions obtained in the simulation of Fig.\ \ref{figDNARes}$a$ for $h=10\ell$ with those obtained in Rosenbluth sampling.\cite{PNAS_2012,Patrick_JCP_2012}
As expected,\cite{Batoulis,Mooij}
we find that the distributions of the static runs\cite{PNAS_2012,Patrick_JCP_2012} are very different from the distributions obtained with tCBMC. 
Moreover two different equilibrium runs (in which jumps between free and bound states were forbidden) provide the same distributions as tCBMC 
(full lines in Fig.\ \ref{figDNARes}).
 This confirms that tCBMC is indeed an equilibrium run. 
 { 
 If this validates tCBMC from a technical perspective, we believe that the strength 
 of the method, as compared to static approaches, lies in its versatility. This is illustrated in the next section. 
 }


\begin{figure}
\begin{center}
\includegraphics[scale=0.26]{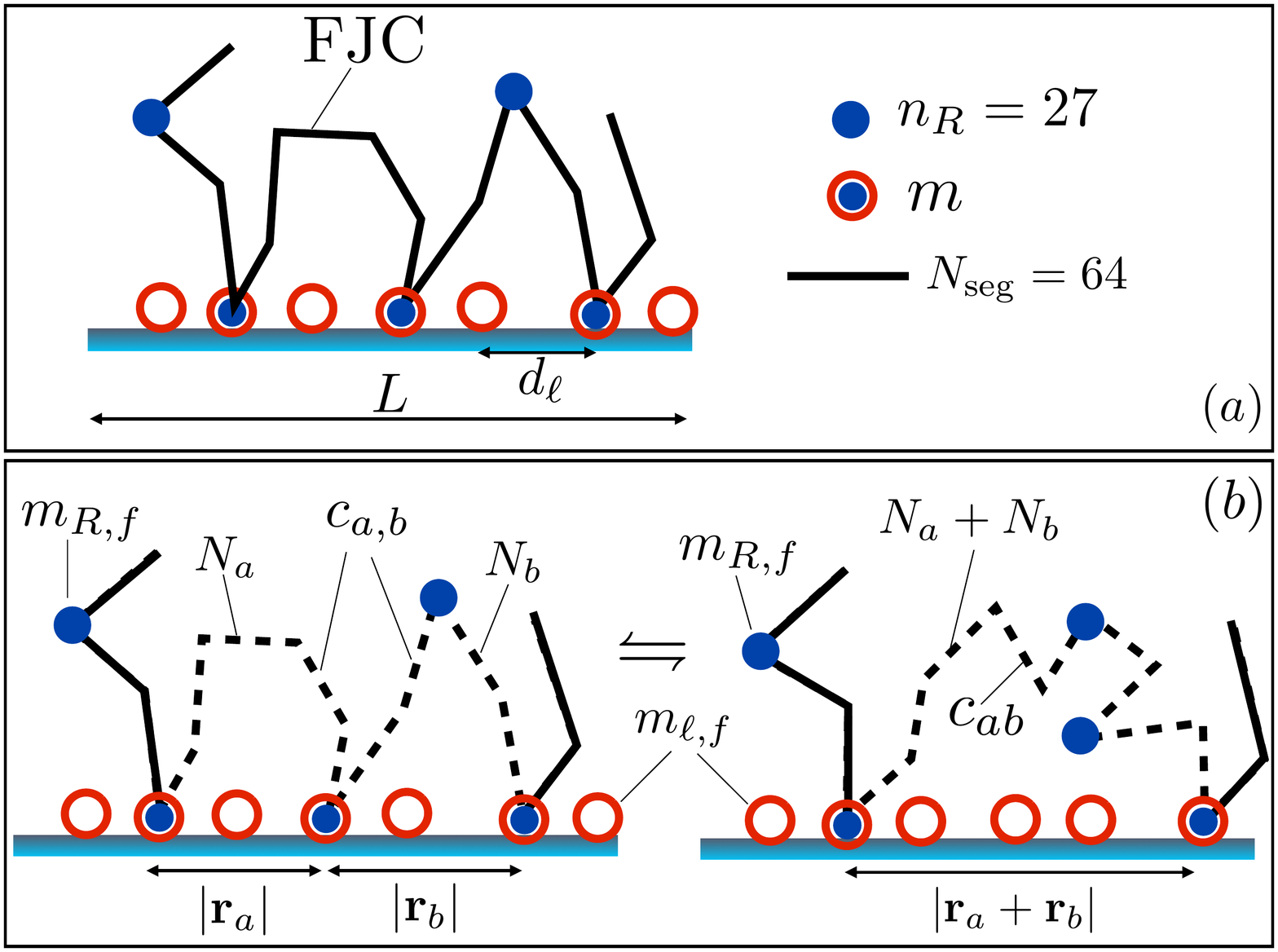}
\end{center}
\caption{\label{figFP} A system of polymers functionalized by receptors targeting a surface decorated by ligands:\cite{Galina} $(a)$ definition of the model parameters, and $(b)$ loop inserting/deleting MC moves.  
}
\end{figure}

\section{Adsorbed polymers}\label{Sec:SE}

In this section we want to demonstrate how tCBMC can handle situations in which the typical configurations 
include a large number of different topologies  
separated by entropic barriers that cannot be easily overcome by standard simulations.
We will study a system of polymers functionalized by receptors targeting a surface decorated by ligands (Fig.\ \ref{figFP}$a$). 
Most of the studies on polymer binding to surface (e.g.\ Ref.\ \onlinecite{Eisenriegler:1982,De-Virgiliis,Daoud:1991,Netz:2003}) have focused on non--selective adsorption { in which each monomer of the chain interacts with every element of the surface}. 
{ 
Here, we consider the case in which a selected fraction of the monomers carries a binding site (receptor) whereas all other monomers cannot bind to the surface. Moreover, the surface is considered to display discrete binding sites (ligands) at a given surface density (Fig.\ \ref{figFP}).
In this case the adsorbed chain can neatly be decomposed into 
a series of loops encompassed by 
two tails (Fig.\ \ref{figFP}). 
The selective--monomer case is usually addressed by standard simulations\cite{SA_num} or using theoretical modelling.\cite{SA_th,SA_dft}
}

In this section we { provide a valuable alternative  by using the algorithm of Sec.\ \ref{Sec:TheMethod} to design } MC moves that allow to create/destroy loops in one go.
We will use this algorithm to calculate the density of states  ($\Omega(m)$) of adsorbed constructs 
which we define as
\begin{eqnarray}
Z_\mathrm{ads} = \sum_{m\ge 1} Z(m) &\qquad&  Z(m) = \Omega(m) \Big[ { e^{-\beta \DG_0} 
\over \rho_\ominus }\Big]^m 
\label{eq:Z}
\end{eqnarray}
where $Z_\mathrm{ads}$ ($Z(m)$) is the partition function of an adsorbed chain (binding $m$ ligands), 
and $\DG_0$ is the free-energy of the receptor--ligand dimerisation.
 We will compare our findings with the results of a mean field theory that will be developed to rationalise recent findings,\cite{Galina} and that has been detailed in Appendix \ref{app:MF}. 
Our methodology has the potential to unveil how the typical configurations of the adsorbed chain affect
 the { thermodynamics of adsorption.\cite{Galina}
 } 
 This is more complicated than the case of 
 functionalized nano--particles for which the configurational costs are simply additive in the number of bound tethers.\cite{Francisco_PNAS_2011}

The system of Fig.\ \ref{figFP}$a$ has been motivated by recent experiments on
constructs of the biological polysaccharide hyaluronan (HA) functionalized with
 hosts reacting with guests immobilised on a surface.\cite{Galina}
HA have an unusually large Kuhn segment length,\cite{Attili:2012} and were here modelled 
by non-interacting FJCs made of $N_\mathrm{seg} = 64$ segments of length $a_K = 14\,$nm.
On a FJC we randomly distribute  $n_R$ receptors placed at the junction points between two segments { (notice we have {$N_\mathrm{seg} + 1$} available spots).} Each receptor can selectively bind a surface on which ligands are randomly distributed at a density equal to $1/d_\ell^2$, where $d_\ell$ is the averaged distance between ligands.  In this study we have used $n_R=27$.\cite{Galina}
 Each topology is characterised by a number $m$ of formed ligand-receptor complexes (Fig.\ \ref{figFP}$a$).
 %
 %

The tCBMC scheme employed is depicted in Fig.\  \ref{figFP}$b$. A loop is generated/destroyed as a result 
of the binding/unbinding of a randomly chosen receptor to/from a randomly chosen ligand.
Such a scheme requires the ability to generate loop configurations with fixed end--points ($c_{ab}$ in Fig.\ \ref{figFP}$b$) and double--loop configurations with three fixed point constraints ($c_{a,b}$ in Fig.\ \ref{figFP}$b$).
Notice that in Fig.\ \ref{figFP}$b$, $c_{ab}$ differs from $c_{a,b}$ only for the dashed parts of the chains. The remaining fraction of the polymer (full lines) is not affected by a single step implementation of the MC move and may include more loops.
We generate configurations of type $c_{a,b}$ and $c_{ab}$ using the same procedure outlined in Sec.\ \ref{Sec:TheMethod}. 
In particular when growing a loop of length $N_a$, we use the FJC end--to--end probability distribution function $p(|{\bf x}_{i-1}+{\bf u}_{i,\alpha}-{\bf r}_a|,N_a-i)$\cite{yamakawa1,yamakawa2} to generate trial vectors that are then sampled using the corresponding Rosenbluth weights (Sec.\ \ref{Sec:TheMethod}). Here ${\bf x}_{i-1}$ is the end point of the $(i-1)$--th segment relative to the starting point of the loop.
With such a procedure we can (re)grow single and double loop configurations and calculate their Rosenbluth weight (that we label by $W^{(a,b)}$ and $W^{(ab)}$ to distinguish between the two different topologies).

Given the previous procedure of generating configurations, the algorithm works as follows.
When making a loop we attempt a reaction between a random receptor (chosen within the $m_{R,f}$ free ones) on the polymer and a random ligand (chosen within the $m_{\ell,f}$ ones that are free) on the surface (see Fig.\ \ref{figFP}).
We then grow a new configuration of type $c^{(n)}_{a,b}$ and calculate the corresponding Rosenbluth weights $W^{(a,b)}(c^{(n)}_{a,b})$. Similarly, we retrace the old configuration $c^{(o)}_{ab}$ and calculate its Rosenbluth weight $W^{(ab)}(c^{(o)}_{ab})$. 
In the reverse move we try to un--bind a host--guest complex randomly chosen within the $m$ that are present in the system (Fig.\ \ref{figFP}$a$). Similarly to what was done before, we grow a new single loop configuration $c^{(n)}_{ab}$, (re)grow the current double-loop configuration ($c^{(o)}_{a,b}$), and measure the corresponding Rosenbluth weight ($W^{(ab)}(c^{(n)}_{ab})$ and $W^{(ab)}(c^{(o)}_{a,b})$). 
The acceptance rules for the two moves are then given by
\begin{eqnarray}
acc_{ab \to a,b}  &=&
\mathrm{min}\Big[ 1,  {m_{R,f} m_{\ell,f} \over m+1}  {e^{-\beta \DG_0}\over \rho_\ominus} 
\cdot  \label{AccPol_1}\\
&& \cdot {W^{(a,b)}(c^{(n)}_{a,b})  \over W^{(ab)}(c^{(o)}_{ab}) } {p(|{\bf r}_a|,N_a) p(|{\bf r}_b|,N_b)  \over p(|{\bf r}_a+{\bf r}_b|,N_a+N_b) }\Big]
\nonumber \\
acc_{a,b \to ab} &=& 
\mathrm{min}\Big[ 1,  { m \over (m_{R,f}+1) (m_{\ell,f}+1) }  {\rho_\ominus \over e^{-\beta \DG_0}} 
\cdot  \label{AccPol_2}\\
&& \cdot {W^{(ab)}(c^{(n)}_{ab})  \over W^{(a,b)}(c^{(o)}_{a,b}) } { p(|{\bf r}_a+{\bf r}_b|,N_a+N_b)  \over p(|{\bf r}_a|,N_a) p(|{\bf r}_b|,N_b)   }\Big] \, .
\nonumber
\end{eqnarray}
We recognise that the structure of Eqs.\ \ref{AccPol_1} and \ref{AccPol_2} is the same as that of Eqs.\ \ref{acceptances} (with $W^{(a,b)}$ and $W^{(ab)}$ replacing $W^{(f)}$ and $W^{(b)}$ respectively). 
The pre--factors are due to the fact that, following the flow of the algorithm, the probability to generate a $c^{(n)}_{a,b}$ or a $c^{(n)}_{ab}$ configuration is equal to $1/(m_{R,f} m_{\ell,f})$ and $1/m$ respectively.

Notice that the randomly selected reacting receptor could be on the tail of the construct (rather than in a loop as in Fig.\ \ref{figFP}$b$). In this case the algorithm should sample between a configuration made of a tail and a configuration made of a tail plus a loop. 
This can be
easily implemented by generalising the way configurations are generated and Eqs.\ \ref{AccPol_1} and \ref{AccPol_2}. For completeness, we detail how the algorithm works in this case in Appendix \ref{AppAcc}.

We also report that more efficient runs were obtained by implementing a 
MC move that re--arranges two loops by swapping the ligand which a receptor is bound to. 
The details are also reported in Appendix \ref{AppAcc}. 
Notice that such a move would be required to guarantee the ergodicity of the algorithm in the case that the choice of the free ligand to bind/unbind (when making/destroying a loop) would be restricted (for efficiency purposes) to a region enclosing the tethering points. 
In that case certain loops with stretched strands could be unreachable by an algorithm that would only use making/destroying loop moves.
{
Alternatively one can think of biasing the choice of the ligands in more subtle ways that also depend on the length of the loops/tails that encompass the randomly selected receptor, as well as on the positions of the ligands to whom such loops/tails are tethered.
}
{Those complications were} avoided in our current study. In particular, we 
randomly distributed ligands across a square of side 
$L=11\cdot a_K$ with periodic boundary conditions and, when attempting to bind/unbind receptors, the ligands were chosen uniformly.

We are now in a position to sample between the different topologies of an adsorbed polymer by adding and removing loops. 
This could be hampered by high free energy barriers resulting in runs exploring only a very few distinct $m$.
%
%
%
To avoid this kind of trapping, we have used the successive umbrella sampling (SUS) scheme of Virnau and M\"uller.\cite{SUS} 
At a certain step of the simulation we only allow sampling between configurations with $m$ and $m-1$ attached receptors. This implies that if the system is in a state with $m$ ($m-1$ receptors) and we attempt to create (destroy) a loop the MC move is immediately rejected. 
By sequentially moving the window within which the sampling is constrained, we can reconstruct $Z(m)$ by using\cite{SUS}  
\begin{eqnarray}
\alpha_m = {\Omega (m) \over \Omega (m-1)}= {N_{m}\over N_{m-1} } 
& \qquad &
{\Omega (m)\over \Omega (1)}  = \prod_{i=2}^m \alpha_m 
 \, ,
 \label{eq:SUS}
\end{eqnarray}
where $N_m$ and $N_{m-1}$ are the number of times that the run has visited a configuration 
with $m$ and $m-1$ complexes formed when constrained between $m-1$, and $m$.

\begin{figure}
\begin{center}
\includegraphics[scale=0.42]{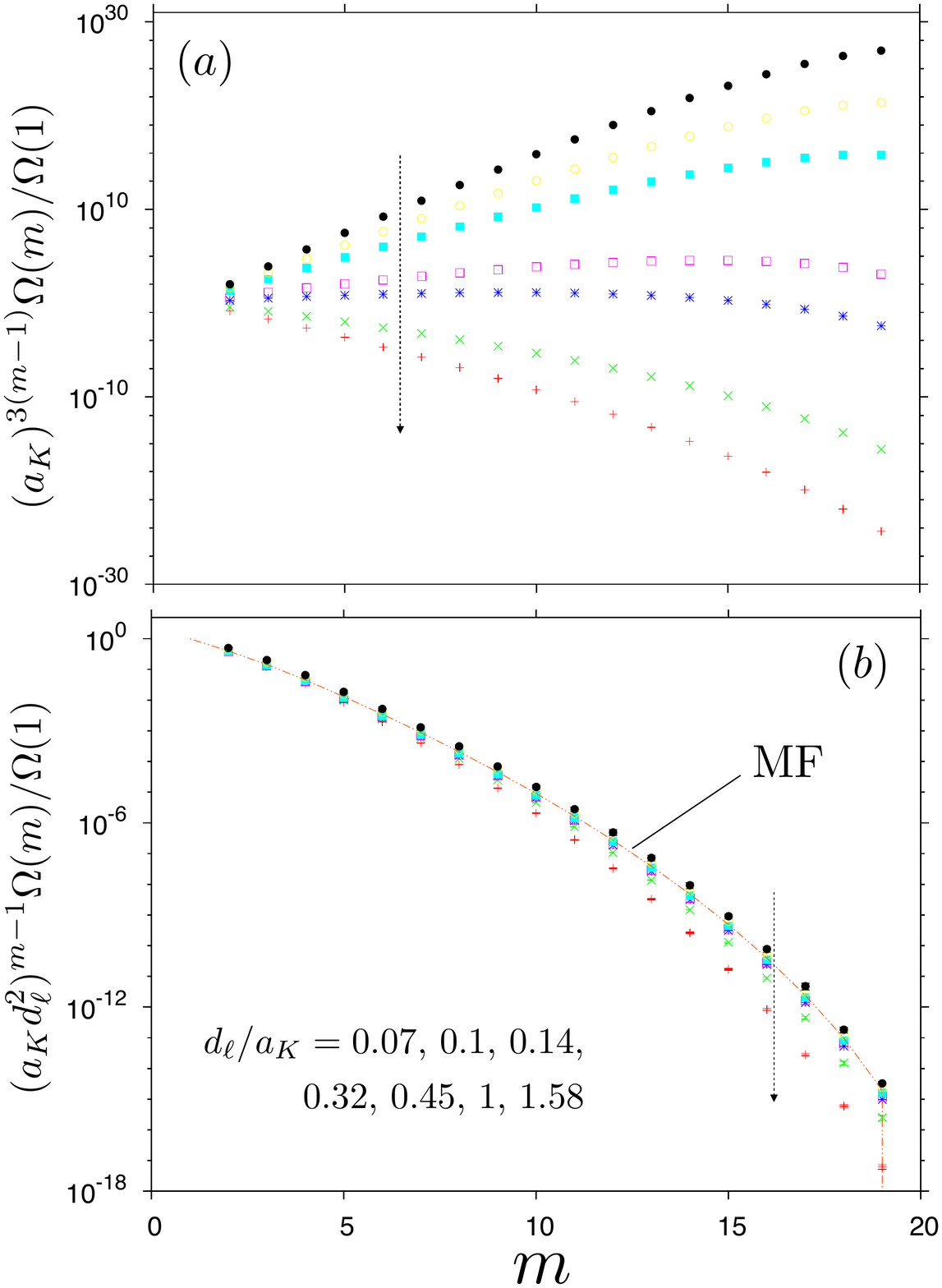}
\end{center}
\vspace{-1.cm}
\caption{\label{figFPRes} 
Density of states {\em versus} number of bound receptors at different ligand concentrations (the arrow direction points toward data sets with higher $d_\ell$). In part $(a)$ $\Omega(m)$ is reported in simulation units while in $(b)$ we test the mean field scaling prediction of Eq.\ \ref{MF:body} (full line). 
}
\end{figure}

In Figure \ref{figFPRes}$a$ we report the density of states $\Omega(m)$ (Eqs.\ \ref{eq:SUS} and \ref{eq:Z}) normalised by $\Omega(1)$.
 While increasing the average ligand--ligand distance $d_\ell$, $\Omega(m)$ decreases. Interestingly, in intermediate regions, $\Omega(m)$ exhibits a maximum.

This finding is easily interpreted. $\Omega(m)$ is the result of the competition between the combinatorial gain of many ligands that can bind many receptors and the configurational cost 
associated to the formation of loops and to the { polymer surface interaction} (Fig.\ \ref{figFP}).\cite{CombConf}
Increasing the ligand density (i.e.\ decreasing $d_\ell$) rises the multivalency of the system, and the density of states increases because of the combinatorial gain.

This statement can be made more rigorous considering a MF theory in which ligands are regularly
distributed with a homogeneous density $1/d_\ell^2$. 
We can show that in this approximation (see Appendix \ref{app:MF}) the partition function can be written as
\begin{eqnarray}
\Omega_\mathrm{MF}(m) &=& 
n_\ell \Big[ {\sqrt{3/2\pi}\over  d_\ell^2 a_K } \Big]^{m-1}  \Psi(m) \, ,
\label{MF:body}
\end{eqnarray}
where $\Psi(m)$ is a function that only depends on the architecture of the functional chain  (see Eq.\ \ref{eq:Psi}), and $n_\ell$ is the number of ligands present on the plane.
In particular $\Psi$ accounts for the multivalency of the receptors on the chain but is independent of $d_\ell$.   
Interestingly Eq.\ \ref{MF:body} predicts a scaling relation between $\Omega$ and $d_\ell$ 
that has been tested in Fig.\ \ref{figFPRes}$b$ for the simulation results of Fig.\ \ref{figFPRes}$a$. 
Satisfactorily, when plotting $\Omega(m)d_\ell^{2( m-1)} $ we find a nice collapse of the density of states at different 
ligand concentrations. 
Importantly at small $d_\ell$ simulations agree with the MF theory.
This is not the case when $d_\ell$ becomes comparable with the length of the Kuhn segment $a_K$.
Indeed in this case the typical loop configurations become more stretched, resulting in a density of states smaller than what is predicted by the MF theory.
Overall, the results of this section 
validate the use of tCBMC to study selective adsorption of polymers, { the key advantage being 
the possibility to sample directly between different adsorbed states by means of dedicated MC 
moves.}

\section{Discussions}\label{Sec:Dis}

Functionalizing complex macromolecules by reactive elements is nowadays a popular tool to 
engineer self--assembling systems and smart aggregates. 
In spite of the high degree of designability of these materials,
efficient simulation methods are hampered by the multi--scale nature of these systems.

In this paper we have developed a Monte Carlo scheme dedicated to the study of thermally reconfigurable supramolecular networks.
This scheme combines an implicit treatment of the reactive sites with coarse grained 
simulations that are used to sample the polymeric network.
Based on our previous works,\cite{PNAS_2012,Patrick_JCP_2012} and similar to what is done in existing literature,\cite{Esco_EP,Vendr_EP,Dij_EP,Siep_EP} we have used schemes that can generate polymers whose configurations are constrained by the reactions of the active spots. 
Comparably to what is done in configurational bias MC,\cite{CBMC} we use the bias measured while 
generating these configurations to implement dynamic MC moves between them. 
The novel development in the present case is that we
 were able to { directly} sample between states with different 
topologies. 
%

First we tested the algorithm by considering tethered constructs tipped by reactive spots as 
in systems of DNA coated colloids. We have demonstrated that the proposed method can reproduce 
the correct hybridisation free energy previously obtained using established methods. 

We have then studied a system of polymers functionalized by receptors binding ligands distributed 
on a surface. 
In this case many possible topologies are present, with polymers exhibiting multiple loops while binding different groups of receptors to different groups of ligands. 
Such topologies are separated by entropic barriers that hamper the efficiency of algorithms based on 
local Monte Carlo moves.
We have 
demonstrated the ability of the proposed algorithm to handle also this system, 
supporting the usefulness of the proposed method with respect to existing techniques.
This has been done by measuring the density of states of adsorbed chains and by comparing them to the predictions of a mean field theory.
{ 
In appendix \ref{app:MF} we show how this quantity can be used to derive, e.g., 
binding isotherms  
and to identify regions in parameter space where the functionalized constructs
discriminate sharply between surfaces with high and low ligand coverage.\cite{Francisco_PNAS_2011,Galina}
This ``superselective'' behaviour is desirable when engineering smart systems for drug delivery.
}

We believe that the proposed method could support the understanding and the design of supramolecular 
systems. For instance, concerning DNA coated colloids, it will allow to calculate the full density of states 
of two particles {cross-linked} by a given number of bridges.  
This will highlight the role played by tether--tether interactions which is usually neglected in the modelling 
of micron-sized particles\cite{Rogers_PNAS,PNAS_2012,Patrick_JCP_2012} but which has been shown to 
be relevant for particles of sub-micron size.\cite{Mladek}
Concerning selective targeting, the proposed method could aid the design of functionalized 
chains resulting in desired  properties. 
In this respect, it will be important to generalise our scheme to {worm-like} chain models and study 
how 
the ligand--receptor affinity is altered when the receptor is mounted on a semi--flexible 
segment.\cite{Hsu:2013}
{
This can be done in view of the fact that
 it is possible to grow fixed end--point chains featuring strong intramolecular interactions
 between adjacent segments of the chains.\cite{Siep_EP}
}
The study of excluded volume interactions between polymers is also desirable.
%
%
In a more general perspective, it will be interesting to explore the usefulness of the method when applied to other relevant systems like, for instance,
 network forming polymers.\cite{Dimopoulos_2010,Biffi_PNAS,Schmid_PRL,Meijer} 
 This will deserve future investigations.

{\bf Acknowledgements.} Computational resources were kindly provided by the HCP Computer Center of the ULB/VUB (Brussels), and by the ``Consortium des Equipements pour le Calcul Intensif'' (CECI). We acknowledge Oleg Borisov for fruitful discussions concerning the mean field theory of adsorbed chains. {T.C. acknowledges support from the Herchel Smith fund. G.V.D. acknowledges Marie Curie Career Integration Grant ÒCELLMULTIVINTÓ (PCIG09-GA-2011-293803).}

\appendix

\section{A Mean Field Theory for Adsorbed Chains}
\label{app:MF}

In this appendix we derive the mean field estimate that has been used in section \ref{Sec:SE} to compare the 
partition function of an adsorbed polymer $Z(m)$ (Eq.\ \ref{MF:body} and Fig.\ \ref{figFPRes}). This is possible 
in view of the fact that we are taking ideal constructs for which the partition function of an adsorbed chain can 
be decomposed into the product of loops and tails.

We first concentrate on calculating the partition function of a tail made of $n$ segments ($Z_\mathrm{tail}(n)$) and the partition function of a loop made of $n$ segments with end points tethered at a distance equal to $r$ ($Z_\mathrm{loop}(n;r)$).  By means of Rosenbluth sampling we obtain 
\begin{eqnarray}
Z_\mathrm{tail}(n)\approx {0.55\over \sqrt{n}} &\qquad \qquad& n\geq 1\, ,
\nonumber \\
Z_\mathrm{loop}(n;r) = {p(r,n)\over n}  &\qquad \qquad & n \geq 2 \, .
\label{TailLoop}
\end{eqnarray}
 $Z_\mathrm{tail}$ and $Z_\mathrm{loop}$ are calculated with respect to the partition function of an ideal 
 chain of length $n$ and fixed starting point.  
In particular, in Eq.\ \ref{TailLoop} the $1/\sqrt{n}$ and $1/n$ terms are the corrections due to the 
impermeability of the plane, while the end point constraint in $Z_\mathrm{loop}$ is accounted 
for by the distribution function $p(r,n)$.\cite{yamakawa1,yamakawa2}  
We define by $x_i$ ($i=1,\cdots , n_R$) the ordered sequence of the positions of the receptors along the chain  ($x_i\in [0,\cdots , N_\mathrm{seg}]$ with $x_i< x_j$ if $i<j$).

Using Eq.\ \ref{TailLoop} we can compute the partition function of a chain binding $m$ ligands (placed in 
${\bf r}_\alpha$, $\alpha=1,\cdots , m$) to the $m$ receptors $x_{\sigma_\alpha}$ 
(where $\sigma_{\alpha}<\sigma_{\beta}$ if $\alpha<\beta$):
\begin{eqnarray}
Z(\{x_{\sigma_\alpha},{\bf r}_\alpha\})&=&Z_\mathrm{tail}(x_{\sigma_1}) Z_\mathrm{tail}(N_\mathrm{seg}-x_{\sigma_m}) \Big[{e^{-\beta \Delta G_0}\over \rho_\ominus }\Big]^m \cdot
\nonumber \\
&& \cdot \prod_{\alpha=2}^m Z_\mathrm{loop}(|{\bf r}_\alpha-{\bf r}_{\alpha-1}|,x_{\sigma_\alpha}-x_{\sigma_{\alpha-1}}).
\nonumber \\
\label{Zmr}
\end{eqnarray}
In the previous expression $\sigma$ labels one of the $n_R!/(m!(n_R-m)!)$ different sets of $m$ receptors taken from the $n_R$ ones present on the chain.

Next we approximate the end--to--end distribution function 
 in Eq.\ \ref{TailLoop} with a 
Gaussian
\begin{eqnarray}
p(r,n) \approx \Big( { 3 \over 2 n a_K^2 \pi} \Big)^{3/2} \exp \Big[-{3 r^2\over 2 a_K^2 n} \Big] \, .
\label{eq:gauss}
\end{eqnarray}
Although this is a good approximation only when $n$ is large we have verified,
 using the explicit form of the end--to--end distance $p(r,n)$,\cite{yamakawa1,yamakawa2}
that the relative discrepancy in the final result is always smaller than $2\%$ (data not shown).

We now approximate the ligand positions  $\{ {\bf r}_i \}$ by a homogeneous distribution of density $1/d_\ell^2$. Practically this allows to replace  sums into integrals as follows
\begin{eqnarray}
\sum_{\{ {\bf r}_\alpha\}} \Big[ \cdot \Big] = \Big[ {1\over d_\ell^2 } \Big]^m \int \mathrm{d} {\bf r}_1 \cdots \mathrm{d} {\bf r}_m \Big[ \cdot \Big] \, .
\label{eq:cont}
\end{eqnarray}

Using Eqs.\ \ref{eq:cont}, \ref{TailLoop}, and \ref{eq:gauss} into Eq.\ 
\ref{Zmr} we can calculate the partition function of a chain adsorbed by the receptors $x_{\sigma_\alpha}$ ($\alpha=1,\cdots,m$)
\begin{eqnarray}
 Z(\{x_{\sigma_\alpha}\} ) \approx
\sum_{\{{\bf r}_i\}} Z(\{x_i, {\bf r}_i\} )
=n_\ell \Big[{e^{-\beta \Delta G_0}\over \rho_\ominus }\Big]^m 
\Big[ {\sqrt{3/2\pi}\over  d_\ell^2 a_K } \Big]^{m-1} &&
\nonumber \\
 \cdot { 0.55^2}
\sqrt{1\over x_{\sigma_1}(N_\mathrm{seg}-x_{\sigma_m}) } 
\prod_{i=2}^m \Big[ {1 \over x_{\sigma_i} -x_{\sigma_{i-1}}} \Big]^{3/2} 
&&
\nonumber \\
\label{eq:qm1}
\end{eqnarray} 
where $n_\ell$ is the number of ligands present on the plane which is taken to be a square of side length 
equal to $11\cdot a_K$ (see Sec.\ \ref{Sec:SE}). 
Finally summing over all the possible sets of $m$ receptors ($\sigma$) we obtain $Z(m)$ as defined in Sec.\ \ref{Sec:SE}:
\begin{eqnarray}
Z(m)&=&  \sum_{\sigma} Z(\{ x_{\sigma_\alpha} \})
\nonumber \\
&=& n_\ell \Big[{e^{-\beta \Delta G_0}\over \rho_\ominus }\Big]^m 
\Big[ {\sqrt{3/2\pi}\over  d_\ell^2 a_K } \Big]^{m-1}  \Psi(m)
\label{eq:Zapp}
\\
\Psi(m) &=& \sum_{\sigma} {0.55^2}
\sqrt{1\over x_{\sigma_1}(N_\mathrm{seg}-x_{\sigma_m}) } 
\prod_{i=2}^m \Big[ {1 \over x_{\sigma_i} -x_{\sigma_{i-1}}} \Big]^{3/2}
\nonumber \\
\label{eq:Psi}
\end{eqnarray}
 $\Psi(m)$ only depends on the position of the receptors on the chain and has been computed by an exact enumeration. In particular for the results of Fig.\ \ref{figFPRes} the 27 receptors were placed at the position $\{x_i\}=\{$2, 3, 5, 7, 10, 14, 15, 19, 23, 29, 31, 34, 35, 36, 39, 44, 48, 50, 51, 54, 55, 56, 57,  59, 60, 61, 62$\}$ along the $N_\mathrm{seg}+1=65$ possible positions. 
 Notice that although we have 27 receptors, In Fig.\ \ref{figFPRes} we never observed more than $m=19$
  complexes reacting. This is due to the fact that receptors that are at the two extremities of the same segment were never allowed to bind simultaneously. Indeed  a concurrent reaction would over--constrain the system. 
Other ways of distributing receptors on the chains (e.g.\ forbidding neighbouring receptors) 
did not alter the general picture.

\begin{figure}
\begin{center}
\includegraphics[scale=0.3]{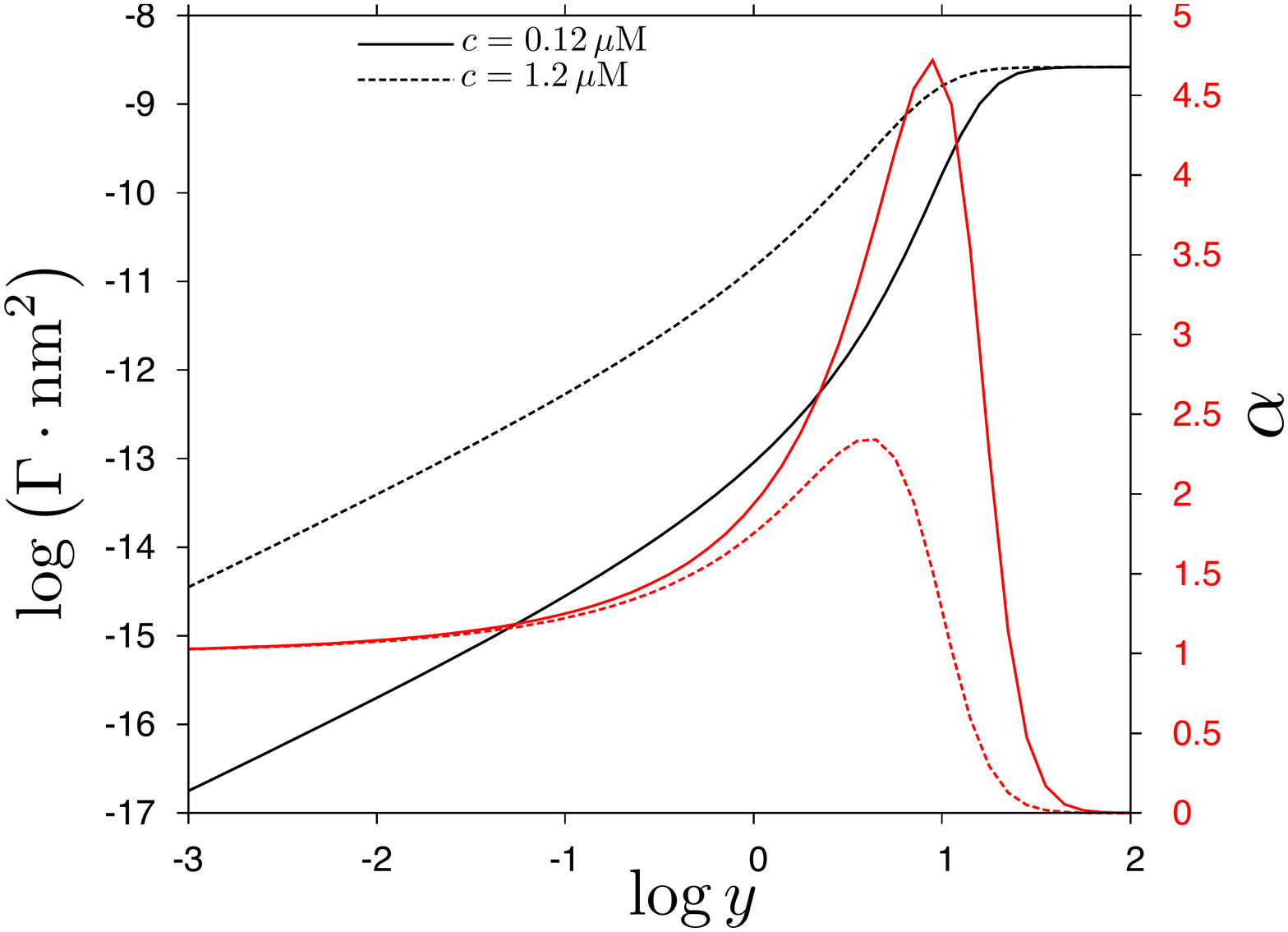}
\end{center}
\caption{\label{figSel} 
{ 
Number of polymers adsorbed {\em per} unit area ($\Gamma$) and selectivity parameter ($\alpha$) as a function of 
the scaling parameter $y$ at two different concentrations.\cite{Galina} 
}
}
\end{figure}

{ 
Using the density of states we can calculate the binding isotherms. In particular, 
following the modelling presented in Ref.\ \onlinecite{Galina}, we divide the functionalized 
plane into square cells of side size equal to $a$. 
Limiting our study to the case in which 
no more than a single polymer can bind a cell, the partition function of a polymer adsorbed onto a 
cell is given by $Z(m)$ (Eq.\ \ref{eq:Zapp}) with 
\begin{eqnarray}
n_\ell = {a^2\over d_\ell^2} \, .
\end{eqnarray}
This scenario is simplified with regard to the real system but has been chosen because 
it is illustrative.
By equalising  the chemical potential of a polymer in the bulk with the chemical potential of an 
adsorbed polymer, we can easily calculate the fraction of cells ($\Theta$) that are occupied by a 
polymer. In particular if we define  
\begin{eqnarray}
Z_\mathrm{bnd} &=& c \sum_m Z(m) = c a^2 a_K  \sum_m y^{m}  \Psi(m)
\label{eq:zbnd}
\end{eqnarray}
where $c$ is the bulk concentration of the polymers and $y$ is the following scaling variable 
\begin{eqnarray}
y &=& { \exp[-\beta \Delta G_0]  \sqrt{3/2 \pi} \over d_\ell^2 a_K \rho_\ominus } 
\label{eq:app_y}
\end{eqnarray}
we find 
\begin{eqnarray}
\Theta = {Z_\mathrm{bnd} \over Z_\mathrm{bnd} +1 } \, .
\end{eqnarray}
Using $\Theta$  we can then derive the number of adsorbed polymers {\em per} unit area 
\begin{eqnarray}
\Gamma = {\Theta \over  a^2 }  \, .
\end{eqnarray}
We notice that the leading term of $\Gamma$ when $y \to 0$ does not depend on $a$ 
($\Gamma \sim c a_K \sum_m y^m \Psi(m)$).

Results for $\Gamma$ are reported in Fig.\ \ref{figSel} (black curves, left $y$-axis), as a function of 
the scaling variable $y$ at two
different polymer concentrations $c$. 
For a given polymer system, the scaling variable $y$ is proportional to the ligand surface density.
Notice that the number of chains adsorbed increases with the scaling variable $y$. 
%
It is useful to calculate  the selectivity parameter defined
as \cite{Francisco_PNAS_2011}
\begin{eqnarray}
\alpha &=& {\mathrm{d} \log \Gamma \over \mathrm{d} \log c_\ell} =  {\mathrm{d} \log \Gamma \over \mathrm{d} \log y}
\end{eqnarray}
where $c_\ell=1/d_\ell^2$ is the density of ligands. 
Notice that $\alpha$ measures how sensible 
the adsorption process is to a change in the ligand surface density.
Calculated values for $\alpha$ are reported in Fig.\ \ref{figSel} (red curves, right $y$-axis).
The superselective region is characterised by  $\alpha>1$ and follows 
previously reported trend.\cite{Francisco_PNAS_2011,Galina}
However, at this point, a quantitative agreement with experiments is still missing.\cite{Galina} 
This is related to limitations of the coarse-grained model for the polymer (that, e.g., neglects
chain rigidity close to a receptor) and to the fact that we did not allow
 multiple polymers binding to the same lattice site in the adsorbed phase. 
These aspects go beyond the scope of this paper and will 
be addressed elsewhere. 

}


\begin{figure}
\begin{center}
\includegraphics[scale=0.26]{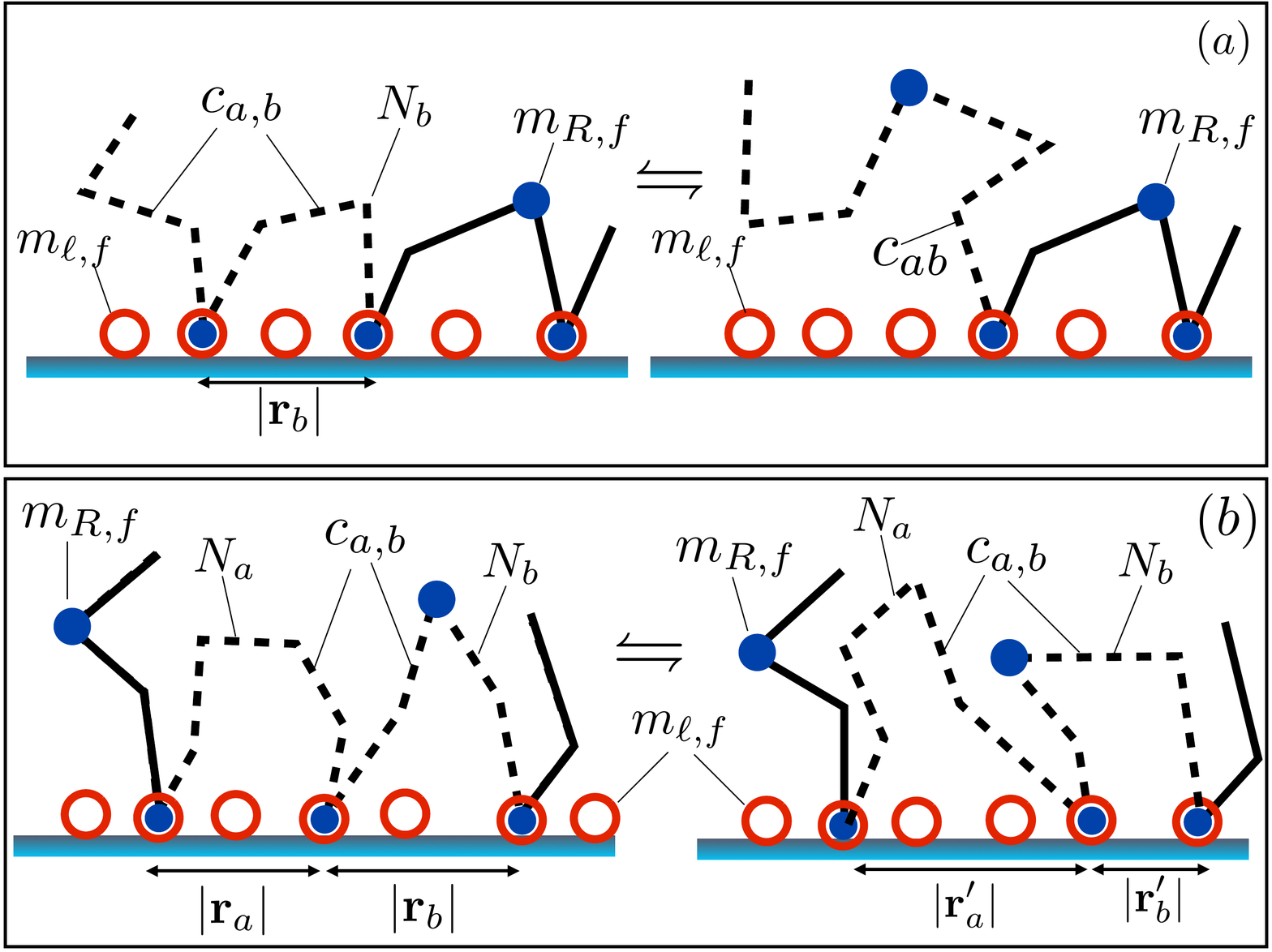}
\end{center}
\caption{\label{figFPApp} 
MC moves by which a receptor on the tail bind/unbind from a ligand $(a)$ and by which a receptor changes ligand to which is bound $(b)$.
}
\end{figure}

\section{Tail Reactions and Ligand Swapping}
\label{AppAcc}
In this section we complete the description of the MC moves introduced in Sec.\ \ref{Sec:SE}.
First we consider the reaction of a receptor on a tail (Fig.\ \ref{figFPApp}$a$). In this case we have to sample between a tail $c_{ab}$ and a tail plus a loop $c_{a,b}$ (dashed lines in Fig.\ \ref{figFPApp}$a$). The generation of  the loop follows what was done in Sec.\ \ref{Sec:SE}. The tail can be generated in the same way as the free constructs in the DNA system (Sec.\ \ref{Sec:DNACC}). In particular the trial segments are not biased by the 
end-to-end distribution function $p$ but are generated with a uniform distribution. 
Using the notation of Sec.\ \ref{Sec:SE}, the acceptance rules are then given by 
\begin{eqnarray}
acc_{ab \to a,b}  &=&
\mathrm{min}\Big[ 1,  {m_{R,f} m_{\ell,f} \over m+1}  {e^{-\beta \DG_0}\over \rho_\ominus} 
\cdot  \label{AccPolApp_1}\\
&& \cdot {W^{(a,b)}(c^{(n)}_{a,b})  \over W^{(ab)}(c^{(o)}_{ab}) }  p(|{\bf r}_b|,N_b)   \Big]
\nonumber \\
acc_{a,b \to ab} &=& 
\mathrm{min}\Big[ 1,  { m \over (m_{R,f}+1) (m_{\ell,f}+1) }  \cdot  \label{AccPolApp_2}\\
&& \cdot  {\rho_\ominus \over e^{-\beta \DG_0}} {W^{(ab)}(c^{(n)}_{ab})  \over W^{(a,b)}(c^{(o)}_{a,b}) } { 1 \over  p(|{\bf r}_b|,N_b)   }\Big] \, .
\nonumber
\end{eqnarray}

We now consider the swapping of a ligand (Fig.\ \ref{figFPApp}$b$). In this case a bound receptor is detached and rebound to another ligand. This implies the construction of double loop configurations ($c_{a,b}$) that is done as described in Sec.\ \ref{Sec:SE}. The acceptance of the receptor displacement is then given by (see Fig.\ \ref{figFPApp})
\begin{eqnarray}
acc_{a,b \to a,b}  &=& 
\mathrm{min}\Big[ 1,  
 {W^{(a,b)}(c^{(n)}_{a,b})  \over W^{(a,b)}(c^{(o)}_{ab}) }   {p(|{\bf r}'_a|,N_a) p(|{\bf r}'_b|,N_b) \over p(|{\bf r}_a|,N_a) p(|{\bf r}_b|,N_b)}   \Big]
 \nonumber \\
\end{eqnarray}
The previous acceptance rule is slightly modified when the receptor that is moved is the tethering point of one of the two tails. In this case we have to sample between configurations made of a loop plus a tail.

%


\begin{thebibliography}{48}
\bibitem [{Note2()}]{optimal_grow}%
  \BibitemOpen
  \bibinfo {note} {There are different ways of doing this. For instance, we can
  think of growing an entire chain before starting with the second one or we
  can grow the chains concurrently, by changing the chain to grow at each step
  of the algorithm. The optimal choice depends on the specific system that is
  investigated.}\BibitemShut {Stop}%
\bibitem [{Note1({\natexlab{b}})}]{CombConf}%
  \BibitemOpen
  \bibinfo {note} {Notice that these contributions are neatly visible in the
  acceptance rules Eqs.\ref{AccPol_1} \ref{AccPol_2}.}\BibitemShut {Stop}%
\end{thebibliography}

\end{document}